\documentclass[pra,aps,showpas,floatfix,twocolumn]{revtex4}
\usepackage{epsfig} \usepackage{graphics} \usepackage{bm} \usepackage{amssymb}
\begin{document}
\title{Quantum dynamics of cavity assisted photoassociation of Bose-Einstein condensed atoms}
\author{Christopher~P.~Search, J. Mauricio Campuzano, and Marko Zivkovic}
\affiliation{Department of Physics and Engineering Physics, Stevens
Institute of Technology, Castle point on the Hudson, Hoboken, NJ
07030, USA}
\date{\today}
\begin{abstract}
We explore the quantum dynamics of photoassociation of Bose-Einstein
condensed atoms into molecules using an optical cavity field. Inside
of an optical resonator, photoassociation of quantum degenerate
atoms involves the interaction of three coupled quantum fields for
the atoms, molecules, and the photons. The feedback created by a
high-Q optical cavity causes the cavity field to become a dynamical
quantity whose behavior is linked in a nonlinear manner to the atoms
inside and where vacuum fluctuations have a more important role than
in free space. We develop and compare several methods for
calculating the dynamics of the atom-molecule conversion process
with a coherently driven cavity field. We first introduce an
alternate operator representation for the Hamiltonian from which we
derive an improved form of mean field theory and an approximate
solution of the Heisenberg-Langevin (HL) equations that properly
accounts for quantum noise in the cavity field. It is shown that our
improved mean field theory corrects several deficiencies in
traditional mean field theory based on expectation values of
annihilation/creation operators. Also, we show by direct comparison
to numerical solutions of the density matrix equations that our
approximate quantum solution of HL equations gives an accurate
description of weakly or undriven cavities where mean field theories
break down.
\end{abstract}
 \maketitle

\section{Introduction}
In the last decade, there has been considerable interest in creating
ultra-cold quantum degenerate molecular systems because of the
potential for improved understanding of molecular physics and
interactions, exploring new types of many-body systems such as
condensates of dipolar molecules and the BCS-BEC crossover, and the
generation of entangled atoms by controlled dissociation of
molecules \cite{search-progress-optics}. Two techniques,
magnetically tunable Feshbach resonances and photoassociation, have
been developed to create ultra-cold molecules that start first from
laser cooled atoms and then induce controlled chemical bonding
between the atoms. Feshbach resonances are the most widely used and
have been successfully applied by numerous research groups to create
molecular dimers starting from either a Bose-Einstein condensate
(BEC) \cite{mol-1,durr,xu-2003,herbig} or a Fermi gas
\cite{regal-2003,strecker-2003,jochim-2003,cubizolles-2003}. This
work culminated in the formation of a molecular Bose-Einstein
condensate (MBEC) \cite{mol-BEC-K,mol-BEC-Li}.

Besides Feshbach resonances, experiments have demonstrated that
two-photon Raman photoassociation can also be used to create
ultra-cold molecules \cite{wynar,julienne1,rom,winkler}. Two-photon
photoassociation has the added benefit that the frequency difference
between the two optical fields can be used to select a particular
rotational-vibrational state including the rotational-vibrational
ground state\cite{tsai,jones}. This gives photoassociation an
advantage over Feshbach resonances since Feshbach molecules are
often very weakly bound in high energy vibrational states that
quickly decay to lower lying vibrational states via inelastic
collisions \cite{xu-2003,yurovsky,mukaiyama}. Recent work has has
used two-photon Raman transitions to create rovibrational ground
state molecules from weakly bound Feshbach molecules
\cite{ni-2008,lang-2008}.

Here we address the problem of two-photon Raman photoassociation of
an atomic BEC inside of an optical cavity that is coherently driven.
Due to the cavity, photons circulate and interact with the atoms and
molecules many times before finally exiting in a manner analogous to
a feedback loop. This feedback amplifies the back action of the
atoms and molecules onto the cavity field causing the light to now
become a dynamical part of the process. In our particular case, we
assume that one of the optical fields used to induce the
atom-molecule conversion is a quantized mode of a driven
Fabrey-Perot resonator while the other field does not correspond to
a cavity mode but is rather a laser directed transverse to the
cavity with sufficient intensity to be treated as a 'classical'
undepleted pump. Our model therefore involves the interaction of
four particles: two atoms are 'destroyed' and a molecule and cavity
photon are 'created' and vice versa. Consequently, the
atom-molecule-cavity photon interaction is analogous to $\chi^{(3)}$
susceptibility in nonlinear optics. This is different from molecule
formation via a Feshbach resonance or free space photoassociation
with undepleted classical lasers where the conversion only involves
two quantum fields: atoms and molecules and is the matter-wave
analog of second harmonic generation of photons with a $\chi^{(2)}$
susceptibility. Coherent photoassociation inside of a cavity
therefore offers the prospect of novel nonlinear dynamics between
the atomic, molecular, and cavity fields and the possibility of
enhanced control over the atom-molecule conversion process.

In an earlier work we analyzed the mean field dynamics and steady
state behavior of cavity assisted photoassociation \cite{markku}
while here we extend that work to study the role of quantum
fluctuations on the dynamics. We analyze the quantum dynamics for
the atomic, molecular, and cavity fields by several methods. First,
an alternate operator representation for the Hamiltonian is
introduced that has an algebra analogous to angular momentum. These
new operators allow us to derive from the Heisenberg-Langevin
equations both an improved form of mean field theory and an
approximate solution for the quantum dynamics that treats the
atom-molecule populations classically but the cavity field and
atom-molecule-photon coherences fully quantum mechanically. The
improved mean field theory incorporates quantum correlations between
the atom and molecules. It also includes a contribution due to
vacuum fluctuations of the atomic and molecular fields that corrects
a deficiency in traditional mean field theory, which fails to
predict atom-molecule Rabi oscillations for resonant transitions
because the solution approaches and becomes stuck at an unstable
equilibrium point. We also compare the improved mean field theory
with our approximate quantum solution and direct integration of the
density matrix equations in the case of small photon and atom
numbers. It is shown that our approximate quantum solution provides
a more accurate description in the case that the cavity driving is
weak or absent in comparison to mean field theories since in this
case the dynamics are initiated by the vacuum fluctuations of the
cavity field.

Before proceeding we note that only a few papers have previously
considered photoassociation inside of a cavity \cite{olsen1,olsen2}.
However, unlike our model, theirs was based on single photon
photoassociation, which is impractical for observing coherent
atom-molecule dynamics because the molecules created are in
electronic excited states and can rapidly decay due to spontaneous
emission. The authors of Ref. \cite{olsen1,olsen2} employed the
positive-P distribution\cite{QuantumNoise} to analyze the quantum
dynamics of the three coupled fields. The positive-P distribution
has a tendency to become numerically unstable for long times in
highly nonlinear quantum optical systems as the stochastic
trajectories begin to sample unphysical regions of phase space and
must be stabilized by proper choice of a stochastic gauge, which is
often difficult to properly determine \cite{drummond}. In fact, it
has been shown that the equations presented in the earlier work
\cite{olsen1,olsen2} are numerically stable only for a limited range
of parameters and that the photon blockade effect predicted in that
work does not in general occur\cite{jaeyoon-thesis}.

The paper is organized as follows: In section II, we present our
model for cavity assisted photoassociation and the density matrix
equations of motion. Additional details for the physical model can
be found in Ref. \cite{markku}. In section III, we derive our
improved mean field theory and an approximate solution of the
Heisenberg-Langevin equations for the dynamics based on a
pseudo-angular momentum operator representation. In section IV, we
compare numerical results for the density matrix, approximate
Heisenberg-Langevin equations, and mean field theories.

\section{Model}

\begin{figure}[ht]
\begin{center}
\epsfig{file=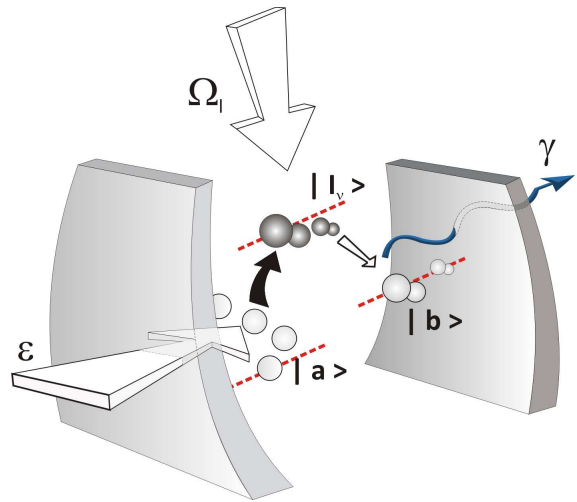,width=7.5cm,} \caption{\label{Fig1} (Color
online) Schematic diagram showing the system under consideration.
The kets $|a\rangle$, $|b\rangle$, and $|I_\nu\rangle$ denote pairs
of atoms, electronic ground state molecules, and electronically
excited molecules, respectively. $\varepsilon$ is the rate at which
the cavity is coherently driven by an external laser while $\gamma$
is the decay rate for photons in the cavity. $\Omega_l$ is the pump
laser that drives the $|a\rangle \rightarrow |I_{\nu}\rangle$
transition.}
\end{center}
\end{figure}

Our starting point is a BEC of atoms inside of an optical cavity, as
depicted in Fig. \ref{Fig1}. The atoms as well as the molecules
formed from them can be trapped inside of the cavity using a far-off
resonant optical trap similar to what has been recently demonstrated
with single atoms in a cavity \cite{boca}. At temperatures $T\approx
0$, we can assume that all of the atoms are in the ground state of
the trapping potential with wave function $\psi_a(r)$. Additionally,
the atoms are assumed to have all been prepared in the same
hyperfine state denoted by $|a\rangle$. Pairs of atoms in
$|a\rangle$ are coupled to electronically excited molecular states
$|I_{\nu}\rangle$, where $\nu$ denotes the vibrational state of the
molecule, via a pump laser with Rabi frequency $\Omega_l$ and
frequency $\omega_l$. The pump is treated as a large amplitude
undepleted source and therefore changes in $\Omega_l$ due to
absorption or stimulated emission are neglected.

The excited molecular states are coupled to molecules in their
electronic and vibrational ground state, $|b \rangle$, via a single
cavity mode. The ground state wave function for the center of mass
of the molecules is denoted by $\psi_b(x)$. Emission of a photon
into the cavity mode takes a molecule from an excited state to its
electronic ground state. Coupling to a single mode can be achieved
by insuring that only a single cavity mode is close to two-photon
resonance for the atom-molecule Raman transition and by positioning
the atoms and molecules around an antinode of the cavity field. The
discrete mode structure of the cavity allows one to select a
particular vibrational state in the electronic ground-state manifold
of the molecules provided the cavity line width, $\gamma$, is less
than the vibrational level spacing, which can be as high as $1$GHz
\cite{tsai}. This would imply a cavity Q-factor of $Q\gg 10^6$,
which has already been achieved with individual atoms trapped inside
of a Fabrey-Perot resonator\cite{boca}. The cavity field frequency
is $\omega_c$ and the vacuum Rabi frequency for the $|I_{\nu}\rangle
\rightarrow |b\rangle$ transition is $g_{cav}$.

The internal energies of states $|I_{\nu}\rangle$ and $|b\rangle$
relative to pairs of atoms in $|a\rangle$ are $\omega_{\nu}$ and
$\Delta\omega<0$, respectively. We assume that the detuning between
the excited states and the pump and cavity fields satisfy,
$\Delta_{\nu}=\omega_{\nu}-\omega_l\approx
(\omega_{\nu}-\Delta\omega)-\omega_c \gg
|g_{cav}|,|\Omega_l|,\gamma_{\nu}$ where $\gamma_{\nu}^{-1}$ is the
lifetime of $|I_{\nu}\rangle$ due to spontaneous emission. Under
these conditions the excited states can be adiabatically eliminated,
leading to two-photon Raman transitions between $|a\rangle$ and
$|b\rangle$ with the resulting Hamiltonian for the
atom-cavity-molecule system,
\begin{equation}
\hat{H}_{am}=\hbar\delta\hat{b}^{\dagger}\hat{b}+i\frac{\hbar
g}{2}(\hat{a}^{\dagger
2}\hat{b}\hat{e}-\hat{a}^{2}\hat{b}^\dagger\hat{e}^\dagger)
+\hbar\chi_a\hat{a}^{\dagger 2} \hat{a}^2+\hbar\chi_b
\hat{b}^{\dagger 2}\hat{b}^2 .\label{HAM}
\end{equation}
Here $\hat{a}$, $\hat{b}$, and $\hat{e}$ are bosonic annihilation
operators for atoms, ground state molecules, and cavity photons,
respectively. Moreover, the molecular and photon operators have been
written in a rotating frame, $\hat{b}\rightarrow
\hat{b}\exp[+i(\omega_c-\omega_l)t]$ and $\hat{e}\rightarrow
\hat{e}\exp[-i\omega_ct]$, to remove all time dependence from the
interaction term. The two-photon detuning is defined as
$\delta=(\Delta\omega-(\omega_{l}-\omega_c))$.

The terms proportional to $\chi_a$ and $\chi_b$ represent the two
body interactions between pairs of atoms and pairs of molecules,
respectively. Interactions involving an atom interacting with a
molecule can be written as
$\chi_{ab}\hat{N}_a\hat{N}_b=\frac{1}{4}\chi_{ab}\left(
\hat{N}^2-\hat{N}_a^2-4\hat{N}_b^2\right)$ where
$\hat{N}=\hat{N}_a+2\hat{N}_b$ is the total number operator and
$\hat{N}_a=\hat{a}^{\dagger}\hat{a}$ and
$\hat{N}_b=\hat{b}^{\dagger}\hat{b}$. Since the total number of
particles is conserved, $[ \hat{H}_{am},\hat{N}]=0$, $\chi_{ab}$ can
be absorbed into a redefinition of $\chi_a$ and $\chi_b$.

The dynamics of the cavity are described by two competing processes.
The first process is cavity decay, which can be treated using the
standard Born-Markov master equation for the density operator
\cite{QuantumNoise},
\begin{equation}
\frac{d\rho}{dt}|_{damping}=\frac{\gamma}{2} \left(
2\hat{e}\rho\hat{e}^{\dagger}-\hat{e}^{\dagger}\hat{e}\rho-
\rho\hat{e}^{\dagger}\hat{e} \right)
\end{equation}
In addition to this, the cavity is coherently driven by an external
laser described by the following Hamiltonian,
\begin{equation}
H_{pump}=i\hbar(\varepsilon\hat{e}^\dagger-\varepsilon^*\hat{e}).
\end{equation}
where we have assumed that the driving laser is resonant with the
cavity mode. The complete equation of motion for the density
operator is then given by,
\begin{equation}
\frac{d\rho}{dt}=\frac{1}{i\hbar}[H_{pump}+H_{am},\rho]+\frac{\gamma}{2}
\left( 2\hat{e}\rho\hat{e}^{\dagger}-\hat{e}^{\dagger}\hat{e}\rho-
\rho\hat{e}^{\dagger}\hat{e} \right) .\label{masterequation}
\end{equation}
The two-photon Rabi frequency is $g=-i\int d^3x \psi_b^*(x)[ g_{cav}
u(x)\Omega_l^*(x)\sum_{\nu}I^*_{a,\nu}I_{b,\nu}/\Delta_{\nu} ]
\psi_a(x)$. $I_{\ell,\nu}$ are the Frank-Condon factors for the
$|\ell=a,b\rangle \rightarrow |I_{\nu}\rangle$ transitions
\cite{drummond-STIRAP} and since typically
$|\Omega_l|/|\Delta_{\nu}|$ and $I_{a,\nu} \ll 1$, $g/\gamma \ll 1$
even if the cavity is in the strong coupling regime $|g_{cav}|>
\gamma$

We represent the density operator in the basis of eigenstates of
$\hat{N}_a$, and $\hat{N}_e=\hat{e}^{\dagger}\hat{e}$,
$|n_a,n_e\rangle$. Because $n_a+2n_b=N$ is a constant of motion, the
molecule number, $n_b$, is completely determined by $n_a$. The
density matrix in this basis, $\langle n'_a,
n'_e|\rho|n_a,n_e\rangle$, is then unwrapped into a column vector,
$\vec{\rho}$, so that Eq. \ref{masterequation} can be written as a
matrix equation, $d\vec{\rho}/dt=\bf{M} \vec{\rho}$, which can now
be integrated using a first order Euler method \cite{savage-1990},
\begin{equation}
\vec{\rho}(t)=\lim_{k\rightarrow \infty} \left( I+(t/k) \bf{M}
\right)^k\vec{\rho}(0) \label{Euler}
\end{equation}
This method has the advantage of using less memory than higher order
ODE solvers but is still limited to small numbers of atoms and
photons. In the next section we develop approximate equations of
motion that incorporate quantum effects but can deal with much
larger experimentally realistic numbers of atoms ($\sim 10^3-10^6$)
and photons.

\section{Pseudo-Angular Momentum Description}
Here we develop a representation of the model developed in the
previous section using pseudo-angular momentum operators that
simplify the form of the Hamiltonian and use this representation to
derive Heisenberg-Langevin equations for the system. We then obtain
mean field equations and approximate equations for the quantum
dynamics from the Heisenberg-Langevin equations.

For an initial state that is an eigenstate of
$\hat{N}=\hat{N}_a+2\hat{N}_b$, the solution of Eq.
\ref{masterequation} will at all later times continue to be an
eigenstate of $\hat{N}$ with the same eigenvalue $N$. In this case,
we can introduce new operators \cite{vardi-anglin}
\begin{eqnarray}
\hat{L}_x&=&\hat{L}_+ + \hat{L}_-=\frac{\sqrt{2}
\left(\hat{a}^{\dagger}\hat{a}^{\dagger}\hat{b}+\hat{b}^{\dagger}\hat{a}\hat{a}
\right) }{N^{3/2}}\\
\hat{L}_y&=&-i(\hat{L}_+ - \hat{L}_-)=\frac{\sqrt{2}
\left(\hat{a}^{\dagger}\hat{a}^{\dagger}\hat{b}-\hat{b}^{\dagger}\hat{a}\hat{a}
\right)}{iN^{3/2}}\\
\hat{L}_z&=&\frac{2\hat{b}^{\dagger}\hat{b}-\hat{a}^{\dagger}\hat{a}}{N}
\end{eqnarray}
and $\hat{L}_{+}^{\dagger}=\hat{L}_{-}$. These operators have the
following commutation relations,
\begin{eqnarray}
\left[\hat{L}_{\pm},\hat{L}_{z} \right]&=&\pm \frac{4}{N}\hat{L}_{\pm} \label{commutator1}\\
\left[\hat{L}_{+},\hat{L}_{-}\right]&=&\frac{3}{2N}\left(
\hat{L}_{z}-1 \right)\left( \hat{L}_{z}+1/3 \right) -\frac{2}{N^2}
\label{commutator2}
\end{eqnarray}
The commutation relations $\left[\hat{L}_{\pm},\hat{L}_{z} \right]$
are of the same form as for angular momentum but the equivalence to
angular momentum is ruined by the commutator
$\left[\hat{L}_{+},\hat{L}_{-}\right]$. We therefore refer to them
as pseudo-angular momentum operators.

Equation \ref{HAM} can be rewritten in terms of these new operators
as
\begin{equation}
\hat{H}_{am}=\frac{\hbar}{4}N\bar{\delta}\hat{L}_{z}+\frac{\hbar}{4}N^2\bar{\chi}\hat{L}_{z}^2
+i\hbar
g\left(\frac{N}{2}\right)^{3/2}\left(\hat{L}_{+}\hat{e}-\hat{L}_{-}\hat{e}^{\dagger}\right)
\end{equation}
where $\bar{\chi}=\chi_a+\chi_b/4$ and
$\bar{\delta}=\delta+N(-2\chi_a(1-1/N)+(\chi_b/2)(1-2/N))$.

It is well known that Born-Markov density matrix equations such as
Eq. \ref{masterequation} are fully equivalent to Heisenberg-Langevin
equations with Markov noise operators in the Heisenberg picture
\cite{QuantumNoise}, which in this case have the form
\begin{eqnarray}
\frac{d}{dt}\hat{L}_{-}&=&i\bar{\delta}\hat{L}_{-}+iN\bar{\chi}\left(\hat{L}_z\hat{L}_{-}+\hat{L}_{-}\hat{L}_{z}\right)+g(2N)^{-1/2}\hat{e}
\nonumber \\
&-&g\left(\frac{3}{4}\right)\left(\frac{N}{2}\right)^{1/2}\left(\hat{L}_{z}-1\right)\left(\hat{L}_{z}+1/3\right)\hat{e}
\label{HL1} \\
\frac{d}{dt}\hat{L}_{z}&=&-g(2N)^{1/2}\left(
\hat{L}_{+}\hat{e}+\hat{L}_{-}\hat{e}^{\dagger}\right) \label{HL2}\\
\frac{d}{dt}\hat{e}&=&\varepsilon-\frac{\gamma}{2}\hat{e}-g\left(\frac{N}{2}\right)^{3/2}\hat{L}_{-}
+\hat{F}(t) \label{HL3}
\end{eqnarray}
Here $\hat{F}(t)$ is a Markov noise operator for the fluctuations of
the electromagnetic reservoir coupled to the cavity mode via the
mirrors with zero mean $\langle \hat{F}(t) \rangle=0$ and the
two-time correlations $\langle \hat{F}(t)\hat{F}(t')\rangle =\langle
\hat{F}^{\dagger}(t)\hat{F}(t')\rangle=0$ and $\langle
\hat{F}(t)\hat{F}^{\dagger}(t')\rangle=\gamma \delta(t-t')$.

Because the Heisenberg-Langevin are nonlinear operator equations,
they cannot be solved exactly. The simplest approximation is that of
mean field theory, which replaces all operators with their c-number
expectation values and factorizes products of operators,
$\hat{L}_{+}\hat{e}\rightarrow \langle \hat{L}_{+}\rangle \langle
\hat{e} \rangle$ thereby ignoring higher order correlations and
non-commutativity of operators. This yields three nonlinear c-number
differential equations,
\begin{eqnarray}
\frac{d}{dt}L_{-}&=&i\bar{\delta} L_{-}+i2N\bar{\chi}L_z
L_{-}+g(2N)^{-1/2}e \nonumber \label{mfl1} \\
&-&g\left(\frac{3}{4}\right)\left(\frac{N}{2}\right)^{1/2}\left(L_{z}-1\right)\left(L_{z}+1/3\right)e
\\
\frac{d}{dt}L_{z}&=&-g(2N)^{1/2}\left(
L_{+}e+L_{-}e^{*}\right) \\
\frac{d}{dt}e&=&\varepsilon-\frac{\gamma}{2}e-g\left(\frac{N}{2}\right)^{3/2}L_{-}
\label{mfl3}
\end{eqnarray}
where the lack of a hat denotes a c-number expectation value,
$A=\langle \hat{A} \rangle$. These equations  are unaffected by
quantum noise since $\langle \hat{F} (t)\rangle=0$ and according to
mean field theory, $\langle \hat{e}\hat{e}^{\dagger}\rangle=\langle
\hat{e}^{\dagger}\hat{e}\rangle=e^*e$.

We note that this form of the mean field approximation is different
from the traditional manifestation with Bose-Einstein condensates or
light where annihilation/creation operators for the fields are
replaced with c-number expectation values. This form was used in our
previous work \cite{markku} and yielded the equations
\begin{eqnarray}
\dot{a}=-2i\chi_a a^{*}a^2+ga^{*}be \label{a-eq} \\
\dot{b}=-i\delta b-2i\chi_b b^2b^{*}-\frac{g}{2}a^2e^* \label{b-eq} \\
\dot{e}=\varepsilon-\frac{\gamma}{2}e-\frac{g}{2}a^2b^* \label{c-eq}
\end{eqnarray}
where $a=\langle \hat{a}\rangle$, $b=\langle \hat{b} \rangle$, and
again $e=\langle \hat{e} \rangle$. In the case that $e(t)$ is
treated as a time independent constant, Eqs. \ref{a-eq} and
\ref{b-eq} are equivalent to earlier mean field studies of
atom-molecule dynamics with Feshbach resonances or photoassociation
\cite{drummond-STIRAP,goral,olsen1,search-progress-optics,gr-jin}.
By contrast Eqs. \ref{mfl1}-\ref{mfl3}, which we refer to as
pseudo-angular momentum mean field theory (PAMMF), represent a
higher order approximation for the atomic and molecular fields than
Eqs. \ref{a-eq}-\ref{c-eq}, which we refer to as amplitude mean
field (AMF) theory. The PAMMF equations automatically incorporate
lowest order quantum correlations between the atomic and molecular
field operators as a consequence of the pseudo-angular momentum
representation, $L_z=\langle
2\hat{b}^{\dagger}\hat{b}-\hat{a}^{\dagger}\hat{a} \rangle /N$ and
$L_-=\sqrt{2} \langle \hat{b}^{\dagger}\hat{a}\hat{a}\rangle
/N^{3/2}$. Additionally, Eq. \ref{mfl1} includes the vacuum
fluctuations of the matter fields resulting from the last term in
the commutator $[\hat{L}_{+},\hat{L}_{-}]$. In the next section we
consider the different predictions for the dynamics made by PAMMF
and AMF.

Mean field theory typically works well for large amplitude quantum
fields. However, initially the cavity field is in the vacuum state
and therefore it cannot be used to describe the initial short time
behavior $\sim \gamma^{-1}$. Additionally, when the cavity is not
driven $\varepsilon=0$, mean field theory completely fails to
describe the dynamics, which are initiated by vacuum fluctuations of
the cavity field. Therefore we develop a solution to Eqs.
\ref{HL1}-\ref{HL3} that properly includes photon noise.

We note that equations \ref{HL1} and \ref{HL3} can be linearized and
solved exactly if one replaces $\hat{L}_z$ with a c-number, $L_z$.
In the case that the c-number is time independent, this approach
would be equivalent to the undepleted pump approximation in
nonlinear quantum optics and the Bogoliubov method for weakly
interacting BEC's. We adopt an approach where the c-number $L_z$ is
instead treated as a dynamical variable whose value is given by the
expectation value of Eq. \ref{HL2}. First we express the equations
for $\hat{L}_-$ and $\hat{e}$ as
\begin{equation}
\frac{d}{dt}\vec{X}=\mathbf{M}\vec{X}+\vec{S} \label{linearizedeqs}
\end{equation}
where $\vec{X}^{T}=\left( \hat{L}_{-}, \hat{e} \right)$ and
$\vec{S}^{T}=\left(0, \varepsilon+\hat{F}(t) \right)$ while
\begin{equation}
\mathbf{M}(t)=\left(
             \begin{array}{cc}
               i\Delta\Omega(t) & -g\beta(t) \\
               -g(N/2)^{3/2} & -\gamma/2 \\
             \end{array}
           \right)
\end{equation}
and $\Delta\Omega=\bar{\delta}+2N\bar{\chi} L_z(t)$ and
$\beta(t)=(3/4)(N/2)^{1/2}(L_z(t)^2-2L_z(t)/3-1/3)-1/\sqrt{2N}$. The
correlation matrix,
\begin{equation}
\mathbf{C}(t)=\langle \vec{X}(t)\vec{X}(t)^{\dagger}\rangle=\left(
                \begin{array}{cc}
                  \langle \hat{L}_-\hat{L}_+\rangle & \langle \hat{L}_-\hat{e}^{\dagger} \rangle\\
                  \langle \hat{e}\hat{L}_+ \rangle & \langle\hat{e}\hat{e}^{\dagger} \rangle \\
                \end{array}
              \right)
\end{equation}
gives us information about the number of cavity photons and the
coherence between the cavity field and the atom-molecule fields,
which can be used to calculate $L_z(t)$. $\mathbf{C}(t)$ has the
equation of motion obtained directly from Eq. \ref{linearizedeqs},
\begin{equation}
\frac{d}{dt}\mathbf{C}=\mathbf{M}\mathbf{C}+\mathbf{C}\mathbf{M}^{\dagger}+\left(
                                                                             \begin{array}{cc}
                                                                               0 & \varepsilon^*\langle \hat{L}_- \rangle \\
                                                                               \varepsilon \langle \hat{L}_-\rangle^* & \gamma+\varepsilon \langle \hat{e}\rangle^*+ \varepsilon^* \langle \hat{e}\rangle\\
                                                                             \end{array}
\right) \label{correlationeqs}
\end{equation}
Equation \ref{correlationeqs} forms a closed set along with the
equations for the expectation value of Eq. \ref{linearizedeqs} and
the expectation value of $\hat{L}_z(t)$,
\begin{equation}
\frac{d}{dt}L_{z}=-g(2N)^{1/2}\left( \langle \hat{e}
\hat{L}_{+}\rangle+\langle\hat{L}_{-}\hat{e}^{\dagger}\rangle\right)
\label{selfconsistentLz}
\end{equation}



We refer to this approach that incorporates the quantum noise of the
cavity field and atom-molecule-photon coherences while treating the
atom-molecule populations classically as the quantum self-consistent
population (QSCP) method. In general this QSCP method is valid for
large $N\gg 1$ in the same manner as the Bogoliubov theory of weakly
interacting condensates \cite{search-progress-optics}. From the
c-number substitution $\hat{L}_z\rightarrow L_z(t)$, one can see
then that the commutation relations Eqs. \ref{commutator1} and
\ref{commutator2} are only preserved in the limit $N\rightarrow
\infty$.  In comparison to AMF and PAMMF equations, Eqs.
\ref{correlationeqs} and \ref{selfconsistentLz} properly account for
quantum noise of the cavity field. The quantum noise of the photons
arise from two sources in the equations for $\mathbf{C}(t)$:
$\langle \hat{e}\hat{e}^{\dagger} \rangle=\langle
\hat{e}^{\dagger}\hat{e} \rangle+1$ representing the vacuum
fluctuations of the cavity mode and the reservoir noise $\langle
\hat{F}(t)\hat{F}^{\dagger}(t')\rangle=\gamma \delta(t-t')$, which
appears as $\gamma$ in the equation for
$\langle\hat{e}\hat{e}^{\dagger} \rangle$.

\section{Numerical Results}
In this section we analyze numerical solutions of the AMF, PAMMF,
QSCP, and exact density matrix equations. In all simulations we use
the initial conditions $N_a(0)=N$ atoms, $N_b(0)=0$ molecules, and
no photons, $N_e(0)=0$.

\subsection{Semiclassical Limit, $N \gg 1$ and $\varepsilon \gg \gamma$}
When the cavity mode is strongly driven, $\varepsilon \gg \gamma$,
the photon noise is negligible and atom-molecule transitions are
dominated by stimulated absorption and emission instead spontaneous
emission into the cavity mode implying that one should be able to
treat the cavity mode classically. Here we compare traditional mean
field theory given by AMF with the improved PAMMF mean field theory
along with the the QSCP method that incorporates cavity noise.

Figure 2 shows that the PAMMF mean field dynamics (Eqs.
\ref{mfl1}-\ref{mfl3}) reproduce qualitatively the behavior of the
the QSCP method (Eqs. \ref{correlationeqs} and
\ref{selfconsistentLz}), which consist of damped Rabi oscillations
that approach the final steady state $N_b=N/2$ molecules and $N_e
\approx 0$ photons. For resonant atom-molecule conversion
($\delta=\chi_a=\chi_b=0$), PAMMF and QSCP agree very closely for
the first few Rabi oscillations while for longer times the QSCP
solutions oscillate at a slightly higher frequency. For off-resonant
transitions ($\delta \neq 0$ or $\chi_a,\chi_b \neq 0$), the
agreement becomes increasingly better and they are indistinguishable
for $|\delta|\gg g\sqrt{N}$. This indicates that quantum
correlations between the cavity field and atom-molecule medium given
by $\langle \hat{e} \hat{L}_{+} \rangle$ play a non-negligible role
in modifying the dynamics when full molecule conversion can occur.

\begin{figure}[ht]
\begin{center}
\epsfig{file=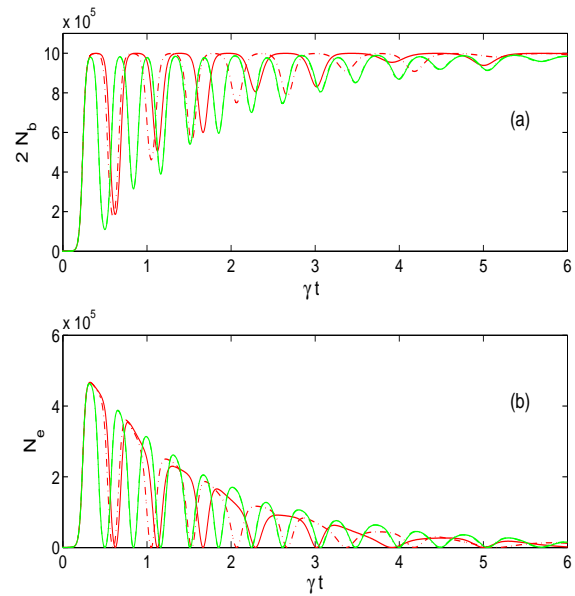,width=1.0\columnwidth,height=8.5cm}
\caption{\label{Fig2} (Color Online) (a) Molecule number and (b)
intracavity photon number calculated using the PAMMF mean field
equations (solid lines) and QSCP equations (dash dot lines) as a
function of time $t$ (in units of $\gamma^{-1}$). In all cases
$N=10^6$, $g/\gamma=5\times 10^{-5}$, and $\varepsilon/\gamma=100$.
Here red lines are for $\delta=\chi_a=\chi_b=0$ while green lines
are for $\delta/\gamma=1$, and $\chi_a=\chi_b=0$. Note that in the
later case (green lines), the two solutions are indistinguishable.}
\end{center}
\end{figure}

By contrast, the AMF solutions (Eqs. \ref{a-eq}-\ref{c-eq}) exhibit
qualitatively very different dynamics from PAMMF and QSCP method in
the resonant case as seen in Fig. 3. In the case
$\delta=\chi_a=\chi_b=0$, AMF equations display no Rabi oscillations
and instead the molecule number grows monotonically until it reaches
the steady state $N_b=N/2$. Atom-molecule oscillations are absent
because as soon as the atomic condensate is fully depleted,
$a(t)=0$, the nonlinear Rabi frequency in Eqs. \ref{a-eq} and
\ref{b-eq} vanishes leading to $\dot{b}=0$ and $\dot{a}=0$. This
behavior is the same as earlier  mean field studies of atom-molecule
conversion via Feshbach resonances or photo-association with
undepleted lasers \cite{goral,vardi-anglin,gr-jin}. In that case the
analytic solution for the molecule number was shown to be
$N_b(t)=(N/2)\tanh^2(\sqrt{2N}|\Omega| t)$ \cite{goral} where
$\Omega$ is atom-molecule coupling, which can be taken to be
equivalent to a time independent $g\langle \hat{e} \rangle /2$ in
our model.

The AMF solution $N_b=N/2$ is an unstable equilibrium
\cite{vardi-anglin} and fluctuations of the atom or molecular fields
would excite the system out of this state leading to atom-molecule
oscillations. The PAMMF equations incorporate the vacuum
fluctuations of the atomic-molecular fields in the term
$g(2N)^{-1/2}e$ in Eq. \ref{mfl1}. Because of this term, even for
$\delta=\chi_a=\chi_b=0$ when there is full conversion into
molecules ($L_z(t)=+1$), one still has $\dot{L}_-=g(2N)^{-1/2}e$.
Figure 4 shows that when $g(2N)^{-1/2}e$ is eliminated from Eq.
\ref{mfl1}, PAMMF and AMF equations produce virtually identical
results \cite{comment}.

Besides the molecule number, the AMF makes different predictions for
the photon number than PAMMF and QSCP. For $\delta=\chi_a=\chi_b=0$,
the PAMMF and QSCP results indicate that the photon number
approaches the steady state $N_e=0$ while the AMF solution predicts
that the photon number grows like $N_e= (2\varepsilon/\gamma)^2$ as
shown in Fig. 5. This behavior is again attributable to the
atom-molecule quantum fluctuations. When the AMF solution reaches
the unstable equilibrium, $N_b=N/2$, the equation for the cavity
field reduces to that of an empty cavity,
$\dot{e}=\varepsilon-\gamma e/2$. By contrast, the PAMMF mean field
equations have the steady state solution $L_z=+1$, $e=0$, and
$L_{-}=(\varepsilon/g)(2/N)^{-3/2}$ for the resonant case.

For nonzero values of $\delta$, $\chi_a$, or $\chi_b$, the AMF
equations do exhibit oscillations between atoms and molecules since
the atom-molecule transition is detuned from perfect resonance,
which prevents full conversion into molecules from occurring. For
example, for small $|\delta|$, PAMMF show higher frequency larger
amplitude oscillations but as $|\delta|$ is increased, the agreement
between AMF and PAMMF becomes increasingly better until the
solutions are again indistinguishable for $|\delta|\gg g\sqrt{N}$ as
seen in Fig. 3.

\begin{figure}[ht]
\begin{center}
\epsfig{file=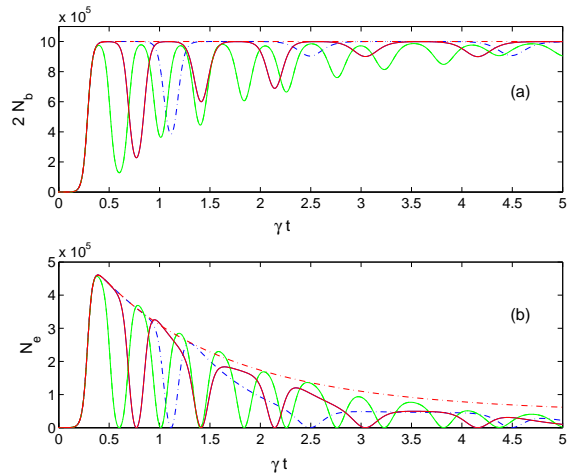,width=1.0\columnwidth,} \caption{\label{Fig3}
(Color Online) (a) Molecule number and (b) intracavity photon
numbers calculated using the PAMMF equations (solid lines) and AMF
equations (dash dot lines) as a function of time $t$ (in units of
$\gamma^{-1}$). In all cases $N=10^6$, $g/\gamma= 4\times 10^{-5}$,
and $\varepsilon/\gamma=100$. Here red lines are
$\delta=\chi_a=\chi_b=0$; blue lines are $\delta/\gamma=0.0001$ and
$\chi_a=\chi_b=0$; green lines are $\delta/\gamma=1$ and
$\chi_a=\chi_b=0$. Note that in the final case (green lines), PAMMF
and AMF are indistinguishable. Also, PAMMF for
$\delta/\gamma=0.0001$ (blue solid line) and PAMMF for $\delta=0$
(red solid line) are indistinguishable.}
\end{center}
\end{figure}

\begin{figure}[ht]
\begin{center}
\epsfig{file=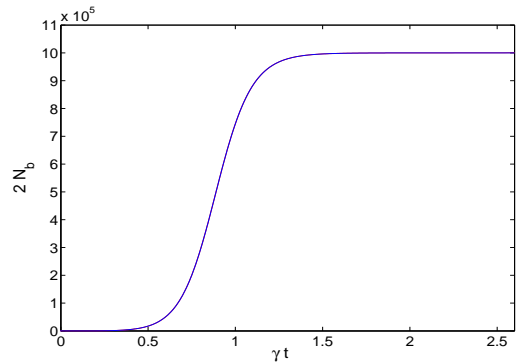,width=0.9\columnwidth,height=5cm}
\caption{\label{Fig3} (Color Online) Molecule number calculated
using the PAMMF equations without the term $g(2N)^{-1/2}e$ in Eq.
\ref{mfl1} (red solid line) and AMF equations (blue solid line).
Here $N=10^6$, $g/\gamma= 10^{-5}$, $\varepsilon/\gamma=100$,
$\delta=\chi_a=\chi_b=0$. The two solutions are indistinguishable.}
\end{center}
\end{figure}

\begin{figure}[ht]
\begin{center}
\epsfig{file=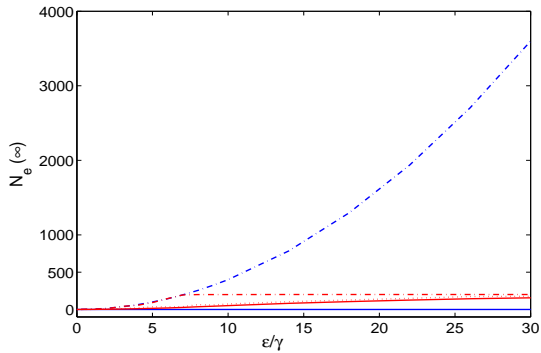,width=0.9\columnwidth,height=5cm}
\caption{\label{Fig3} (Color Online) Steady state photon number as a
function of $\varepsilon/\gamma$ calculated using the QSCP equations
(solid lines), PAMMF equations (dotted lines), and AMF equations
(dashed dot lines). Here $N=10^6$ and $g/\gamma= 5\times 10^{-5}$.
Blue lines are for $\delta=\chi_a=\chi_b=0$ and red lines are
$\delta/\gamma=1$ and $\chi_a=\chi_b=0$.}
\end{center}
\end{figure}

These results indicate that for off-resonant atom-molecule
conversion, all three methods agree. For resonant transitions, the
PAMMF and QSCP show atom-molecule Rabi oscillations due to the
inclusion of the matter field vacuum fluctuations while the QSCP
additionally includes atom-molecule-photon correlations that modify
the effective Rabi frequency.

\subsection{Weakly Driven Cavity, $\varepsilon\ll \gamma$}
For weak driving, $\varepsilon \ll \gamma$, reservoir noise and
cavity vacuum fluctuations are both much stronger than the coherent
driving. In this case, AMF and PAMMF equations are expected to give
an inaccurate description. In fact, for the limiting case
$\varepsilon=0$, both mean field theories predict no dynamics at all
since in the absence of photon fluctuations there is nothing to
initiate the atom-molecule conversion.

First we show a comparison of the PAMMF and QSCP equations for $N\gg
1$ and $\varepsilon \ll \gamma$ in Fig. 6. As one can see for the
QSCP, the conversion of atoms into molecules starts much earlier.
This is easily understood if one solves the QSCP equations for
$L_z(t)$ using perturbation theory. Solving for $L_z(t)$ for short
times from the initial condition $L_z(0)=-1$, one finds that
\begin{equation}
L_z(t)\approx -1+(tg\sqrt{N})^2(1-1/N)(1-\gamma t/6),
\end{equation}
which is correct to order $t^3$ even for arbitrary $\varepsilon$.
The cavity driving $\varepsilon$ only contributes to $L_z(t)$ at
order $t^4$ in perturbation theory with the term
\begin{equation}
L_z(t)|_{\varepsilon}= (g\sqrt{N})^2(1-1/N)|\varepsilon|^2 t^4/4.
\end{equation}
From which we see that the cavity fluctuations dominate the short
time behavior and it is only for later times that the atoms feel the
affect of the cavity driving.

\begin{figure}[ht]
\begin{center}
\epsfig{file=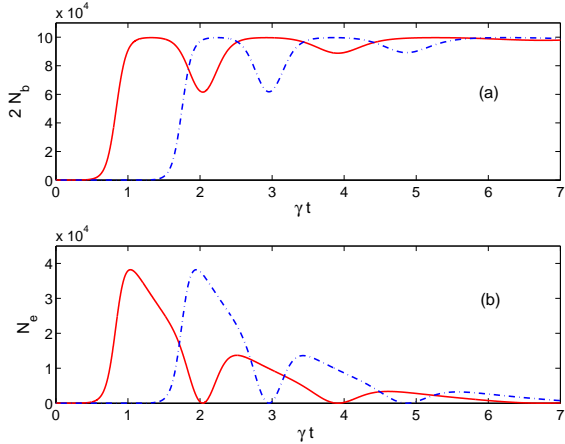,width=1.0\columnwidth,} \caption{ (Color
Online) (a) Molecule number and (b) intracavity photon numbers
calculated using the PAMMF equations (blue dash dot lines) and QSCP
equations (red solid lines) as a function of time $t\gamma$ for the
parameters $N=10^5$, $g/\gamma= 1.5\times 10^{-4}$,
$\delta=\chi_a=\chi_b=0$ and $\varepsilon/\gamma=0.01$.}
\end{center}
\end{figure}

In Figs. 7 and 8 we show solutions of the density matrix (Eq.
\ref{Euler}), QSCP, and PAMMF solutions for initial atom numbers
$N_a=64$ and $\varepsilon=0$ and $0.1\gamma$. One can see that for
times long enough for nearly full conversion to occur, the QSCP and
density matrix solutions show very good agreement while the PAMMF
either predicts no dynamics ($\varepsilon=0$) or dynamics that start
significantly later ($\varepsilon=0.1\gamma$). The molecule number,
$N_b$, for both the PAMMF and QSCP solution approach a steady value
that is below that of the density matrix, $N_b=N/2$. This is a
consequence of the c-number approximation for $\hat{L}_z$, which is
only valid for $N\gg 1$. Our simulations indicate that this error
vanishes as $N\rightarrow \infty$ and both PAMMF and QSCP indicate
full molecule conversion on resonance. Unfortunately, our solutions
of Eq. \ref{Euler} are limited to a maximum of 64 atoms and photons.

\begin{figure}[ht]
\begin{center}
\epsfig{file=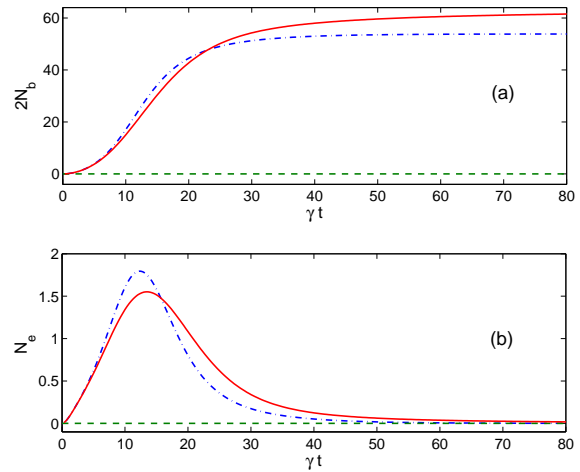,width=1.0\columnwidth,} \caption{ (Color
Online) (a) Molecule number and (b) intracavity photon numbers
calculated using the density matrix (Eq. 5) (red solid lines), PAMMF
equations (green dashed lines), and QSCP equations (blue dashed dot
lines) as a function of time $t\gamma$ for the parameters $N=64$,
$g/\gamma= 0.01$, $\delta=\chi_a=\chi_b=0$ and
$\varepsilon/\gamma=0$.}
\end{center}
\end{figure}

\begin{figure}[ht]
\begin{center}
\epsfig{file=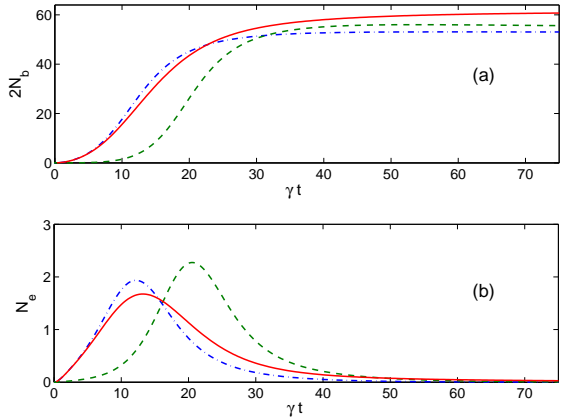,width=1.0\columnwidth,} \caption{ (Color
Online) (a) Molecule number and (b) intracavity photon numbers
calculated using the density matrix (Eq. 5) (red solid lines), PAMMF
equations (green dashed lines), and QSCP equations (blue dashed dot
lines) as a function of time $t\gamma$ for the parameters $N=64$,
$g/\gamma= 0.01$, $\delta=\chi_a=\chi_b=0$ and
$\varepsilon/\gamma=0.1$.}
\end{center}
\end{figure}


\section{Conclusions}
Here we have compared several different approaches for analyzing the
dynamics of intracavity photoassociation. It has been shown that
traditional mean field theory, which replaces annihilation/creation
operators with c-numbers, fails to provide an accurate description
of the dynamics since it ignores the quantum fluctuations of both
the matter fields and the cavity field. We developed two new
approaches that go beyond traditional mean field theory by
introducing a different operator representation for the Hamiltonian
similar to angular momentum. When mean field theory is applied to
the new operator representation, the resulting equations properly
incorporate vacuum fluctuations of the matter fields, which are
necessary for atom-molecule Rabi oscillations to occur. The second
approach, QSCP, which goes one step further by also incorporating
quantum fluctuations and reservoir noise of the cavity field,
provides an accurate description of the short time behavior that
agrees with exact numerical solutions of the density matrix
equations. This final method that incorporates photon noise works
even when the cavity in not driven, which is when both mean field
theories completely fail. Even more important, the QSCP equations
can be easily solved numerically for arbitrary numbers of atoms and
photons while the density matrix can only be solved for relatively
small numbers of atoms and photons, typically less than a 100 each.

In a future work we plan to improve upon the QSCP method in a manner
that will allow us calculate both squeezing and number fluctuations
of the fields.

The work was supported by the National Science Foundation award no.
0757933.


\begin{thebibliography}{99}

\bibitem{search-progress-optics} C. P. Search and P. Meystre,
{\em Nonlinear and quantum optics of atomic and molecular fields,}
in Progress in Optics, Vol. 47, p. 139-214, edited by E. Wolf
(Elsevier, Amsterdam, 2005).

\bibitem{mol-1} E. A. Donley, N. R. Claussen, S. T. Thompson, and C. E.
Wieman, Nature (London) {\bf 417}, 529 (2002).

\bibitem{durr} S. Durr, T. Volz, A. Marte, and G. Rempe, Phys. Rev. Lett. {\bf 92},
020406 (2004).

\bibitem{xu-2003} K. Xu, T. Mukaiyama, J. R. Abo-Shaeer, J. K. Chin, D. E.
Miller, and W. Ketterle, Phys. Rev. Lett. {\bf 91}, 210402 (2003).

\bibitem{herbig} Jens Herbig, Tobias Kraemer, Michael Mark, Tino Weber, Cheng
Chin, Hanns-Christoph N{\"a}gerl, Rudolf Grimm, Science {\bf 301},
1510 (2003).

\bibitem{regal-2003} C. A. Regal, C. Ticknor, J. L. Bohn, and D. S. Jin, Nature
(London) 424, 47 (2003)

\bibitem{strecker-2003} K. E. Strecker, G. B. Partridge, and R. G. Hulet, Phys. Rev.
Lett. 91, 080406 (2003).

\bibitem{jochim-2003} S. Jochim et al., Phys. Rev. Lett. 91, 240402 (2003).

\bibitem{cubizolles-2003} J. Cubizolles et al., Phys. Rev. Lett. 91, 240401 (2003).

\bibitem{mol-BEC-K} M. Greiner, C. A. Regal, and D. S. Jin, Nature (London) 426,
537 (2003).

\bibitem{mol-BEC-Li} M. W. Zwierlein, C. A. Stan, C. H. Schunck, S. M. F. Raupach,
S. Gupta, Z. Hadzibabic, W. Ketterle, Phys. Rev. Lett. 91, 250401
(2003);  S. Jochim, M. Bartenstein, A. Altmeyer, G. Hendl, S. Riedl,
C. Chin, J. Hecker Denschlag, Science 302, 2101 (2003).

\bibitem{wynar} R. Wynar R.S. Freeland, D.J. Han, C. Ryu, D.J. Heinzen, Science {\bf 287}, 1016 (2000);

\bibitem{julienne1} P. S. Julienne, K. Burnett, Y. B. Band, and W. C. Stwalley,
Phys. Rev. A {\bf 58}, R797 (1998); D. Heinzen, R. Wynar, P.
Drummond, and K. Kheruntsyan, Phys. Rev. Lett. {\bf 84}, 5029
(2000).

\bibitem{rom} Tim Rom, Thorsten Best, Mandel, Artur Widera, Markus Greiner, Theodor W. H{\"a}nsch, and Immanuel
Bloch, Phys. Rev. Lett. {\bf 93}, 073002 (2004).

\bibitem{winkler} K. Winkler, G. Thalhammer, M. Theis, H. Ritsch, R. Grimm, and J. Hecker
Denschlag, Phys. Rev. Lett. 95, 063202 (2005).

\bibitem{tsai} C. C. Tsai R. S. Freeland, J. M. Vogels, H. M. J. M. Boesten, B. J. Verhaar, and D. J. Heinzen, Phys. Rev. Lett. {\bf 79}, 1245
(1997).

\bibitem{jones} Kevin M. Jones, Eite Tiesinga, Paul D. Lett, and Paul S.
Julienne, Rev. Mod. Phys. {\bf 78}, 483 (2006).

\bibitem{yurovsky} V. A. Yurovsky, A. Ben-Reuven, P. S. Julienne,
and C. J. Williams, Phys. Rev. A {\bf 60}, R765 (1999); V. A.
Yurovsky and A. Ben-Reuven, Phys. Rev. A {\bf 72}, 053618 (2005).

\bibitem{mukaiyama} T. Mukaiyama, J. R. Abo-Shaeer, K. Xu, J. K.
Chin, and W. Ketterle, Phys. Rev. Lett. {\bf 92}, 180402 (2004).

\bibitem{ni-2008} K. K. Ni, S. Ospelkaus, M. H. G. de Miranda, A.
Pe'er, B. Neyenhuis, J. J. Zirbel, S. Kotochigova, P. S. Julienne,
D. S. Jin, and J. Ye, Science {\bf 322}, 231 (2008).

\bibitem{lang-2008} F. Lang, K. Winkler, C. Strauss, R. Grimm, and
J. Hecker Denschlag, Phys. Rev. Lett. {\bf 101}, 133005 (2008).

\bibitem{markku} Markku Jaaskelainen, Jaeyoon Jeong, and Christopher
P. Search, Phys. Rev. A {\bf 76}, 063615 (2007).

\bibitem{olsen1} M. K. Olsen, J. J. Hope, and L. I. Plimak, Phys. Rev. A 64,
013601 (2001).

\bibitem{olsen2} M. K. Olsen, L. I. Plimak, and M. J. Collett, Phys. Rev. A 64,
063601 (2001).

\bibitem{QuantumNoise} C. W. Gardiner and P. Zoller, "Quantum Noise: A Handbook of Markovian
and Non-Markovian Quantum Stochastic Methods with Applications to
Quantum Optics" (Springer-Verlag, Berlin, 2004).

\bibitem{drummond} M. R. Dowling, M. J. Davis, P. D. Drummond, and J. F.
Corney, J. Comp. Phys. {\bf 220}, 549 (2007); P. Deuar and P. D.
Drummond, J. Phys. A {\bf 39}, 2723 (2006).

\bibitem{jaeyoon-thesis} Jaeyoon Jeong, Ph.D. thesis, Stevens
Institute of Technology, 2007.

\bibitem{boca} A. Boca, R. Miller, K. M. Birnbaum, A. D. Boozer, J. McKeever,
and H. J. Kimble, Phys. Rev. Lett. {\bf 93}, 233603 (2004).



\bibitem{drummond-STIRAP} P. D. Drummond and K. V. Kheruntsyan, D. J. Heinzen and R. H.
Wynar, Phys. Rev. A {\bf 65}, 063619 (2002).

\bibitem{savage-1990} C. M. Savage, Journal of Modern Optics {\bf 37}, 1711
(1990).

\bibitem{vardi-anglin} A. Vardi, V. A. Yurovsky, and J. R. Anglin,
Phys. Rev. A {\bf 64}, 063611 (2001).

\bibitem{goral} Krzysztof Goral, Mariusz Gajda, and Kazimierz
Rzazewski, Phys. Rev. Lett. {\bf 86}, 1397 (2001).

\bibitem{gr-jin} Guang-Ri Jin, Chul Koo Kim, and Kyun Nahm, Phys.
Rev. A {\bf 72}, 045602 (2005).

\bibitem{comment} The authors of Ref. 31 also used the same
pseudo-angular momentum representation for their model of
atom-molecule conversion via a feshbach resonance/photoassociation
but they eliminated the vacuum fluctuation term $\Omega\sqrt{2/N}$
from their equations arguing that since it scaled like $1/\sqrt{N}$,
it was unimportant for $N\gg 1$. As a result, their solutions didn't
display atom-molecule Rabi oscillations.

\end{thebibliography}
\end{document}